\documentclass[12pt]{article}
\usepackage{amsthm,amsmath,natbib}
\usepackage{epsfig}
\usepackage{graphicx}
\usepackage{color}
\usepackage{amsfonts}
\font\tenopen=msbm10
\font\sevenopen=msbm7
\font\fiveopen=msbm5
\newfam\openfam
\def\open{\fam\openfam\tenopen}
\textfont\openfam=\tenopen
\scriptfont\openfam=\sevenopen
\scriptscriptfont\openfam=\fiveopen
\font\title=cmbx10 scaled\magstep1

\def\II{I\negthinspace I}
\def\1I{1{\hskip -3 pt}\hbox{I}}

\def\Z{{\open \mathbb{Z}}}

\def\E{{\open \mathbb{E}}}

\def\II{I\negthinspace I}

\theoremstyle{remark}
\theoremstyle{lemma}
\theoremstyle{definition}
\theoremstyle{corol}
\theoremstyle{proposition}
\theoremstyle{condition}
\theoremstyle{conjecture}
\newtheorem{theorem}{\bf{Theorem}}

\begin{document}

\centerline{\large Testing for Parallelism between Trends in Multiple Time Series}

\bigskip

\centerline{David Degras$^1$, Zhiwei Xu$^2$, Ting Zhang$^1$ and
Wei Biao Wu$^1$}

\centerline{\it $^1$University of Chicago and $\,\,^2$University
of Michigan}

\centerline{May 11, 2011}

\begin{abstract}
This paper considers the inference of trends in multiple,
nonstationary time series. To test whether trends are parallel to
each other, we use a parallelism index based on the
$L^2$-distances between nonparametric trend estimators and their
average. A central limit theorem is obtained for the test statistic
and the test's consistency is established. We propose a
simulation-based approximation to the distribution of the test
statistic, which significantly improves upon the normal approximation.
The test is also applied to devise a clustering algorithm.
Finally, the finite-sample properties of the test are assessed through
simulations and the test methodology is illustrated with
time series from Motorola cell phone activity in the
United States.
\end{abstract}


\section{Introduction}

Comparison of trends or regression curves is a  common problem in
applied sciences. For example in longitudinal clinical studies,
evaluators are interested in comparing response curves for
treatment and control groups. In agriculture, it may be relevant
to compare  at different spatial locations the relationship
between yield per plant and plant density (Young and Bowman,
1995). In biology, assessing parallelism between sets of
dose-response data allows to determine if the biological response
to two substances is similar or if two different biological
environments give similar dose-response curves to the same
substance (Gottschalk and Dunn, 2005). Also in economics, a
standard problem is to compare the yield over time of US Treasury
bills at different maturities, or the evolution of long-term rates
in several countries (Park et al., 2009).

The statistical methodology developed in this paper is motivated
by a collection of time series of cell phone download activity
(applications, audio, images, ringtones, and wall papers)
collected by Motorola in the United States between September 2005
and June 2006. The measurements were collected hourly and
aggregated at the area code level (129 area codes were observed in
total). A question of considerable 
interest is to
determine whether the download trends in the area codes are
identical up to scale differences. (Scale differences can be
expected because of the differences in numbers of phone users for
each area code.) If this hypothesis was true,  it could be
asserted that for those area codes that display slower growth
rates than average, their growth deficit is
non-structural. By placing more advertising efforts and commercial
incentives in these areas, the phone company and its commercial
partners could thus expect cell phone downloads to  increase.
Another interesting application of comparing trends in cell phone
activity pertains to the allocation of bandwidth in phone
networks.

After a pilot study revealed a multiplicative structure in the
trend, seasonality, and irregularities of the time series, a
logarithmic transform was applied to the data so as to stabilize
the variance and obtain an additive (signal + noise) model. In
this context, 
an efficient way to test for proportionality between the trends in
the initial data is to consider the
alternative problem of testing for parallelism between the trends
in the log-transformed time series. From here on, we consider an
additive nonparametric time series model that, in our analysis of
the Motorola data, pertains to the log-transformed data. Suppose
that we observe $N$ time series $\{X_{i t}\}_{t = 1}^T$, $i = 1,
\ldots, N$, according to the model
\begin{eqnarray}\label{eq:Xn}
X_{i t} = \mu_i(t/T) + e_{i t}, \quad
 t = 1, \ldots, T,
\end{eqnarray}
where the $\mu_i$ are unknown smooth regression functions
defined over $[0,1]$ and the $\{e_{i t}\}_{t = 1}^T$ are mean zero
error processes. The scaling device $t/T$ in (\ref{eq:Xn})
indicates that the means $\E X_{i t}$ change smoothly in time, due
to the smoothness of $\mu_i(\cdot)$. It is widely
used in statistics and econometrics; see for example Orbe et al. (2005) and Wu and Zhao (2007). We are interested
in testing whether the $\mu_i$, $i = 1, \ldots, N$, are parallel,
namely, whether there exists a function $\mu$ and numbers  $c_i$
such that
\begin{eqnarray}\label{eq:Null}
H_0: \quad \mu_i(u) =   c_i + \mu(u),
\quad  i=1, \ldots, N, \:\: u\in [0,1].
\end{eqnarray}
Under $H_0$, the $c_i$ represent the vertical shifts between the
curves $\mu_i$ and the reference curve $\mu$. They can be viewed
as nuisance parameters for testing purposes. Note that testing for
parallelism is closely related to testing for equality as, on the
one hand, $H_0$ is formally equivalent to equality of the $N$
centered functions $(\mu_i - \int_0^1 \mu_i(u)du)$ and, on the
other hand, the scalars $\int_0^1 \mu_i(u)du$ can be considered as
known since they are estimated at parametric rates while the
functions $\mu_i$ and $\mu$ are estimated at slower nonparametric
rates.

Various tests for comparing mean functions can be found  in the
regression literature. H\"ardle and Marron (1990) compare two
nonparametric regression curves by testing whether one of them is
a parametric transformation of the other. To test equality of
$N=2$ regression curves in the setup of independent errors, Hall
and Hart (1990) propose a bootstrap test, King et al. (1991) devise a procedure based on the $L^2$-distance between
kernel regression estimators, and Guo and Oyet (2009) apply a
wavelet-based method. For $N \ge 2$, assuming independent errors,
Munk and Dette (1998) use a test based on weighted $L^2$-distances
that requires no smoothing parameter selection. To test whether a
nonparametric mean curve has a certain parametric shape, Bissantz
et al. (2005) and Pawlak and Stadtm\"uller (2007) appeal to signal
processing theory and the Whittaker-Shannon sampling theorem under
independent errors while Degras (2010) utilizes approximate
simultaneous confidence bands in the functional data setup. Under
model and design conditions, their tests can be adapted to assess
parallelism for two mean curves. Young and Bowman (1995) build
ANOVA-type tests for equality and parallelism in $k\ge 2$\break
regression curves under i.i.d. errors. In the time series setup,
to infer equality of two trends, Park et al. (2009) apply a
graphical device assuming stationary, weakly correlated errors; Li
(2006) builds a test based on the cumulative regression functions,
assuming long-memory moving average errors; Fan and Lin (1998) use
an adaptive Neyman test with stationary Gaussian linear error
processes. For random designs of observations,  contributions  to
the comparison of regression curves include Delgado (1993), Koul
and Schick (1997), and Lavergne (2001).

The present work  brings several contributions to the statistical
problem of testing parallelism between trends in multiple time
series. First, studies to date are based on one or both of the
following assumptions: (i) the error processes $\{e_{i t}\}_{t =
1}^T$ in \eqref{eq:Xn} are independent in time, or more generally
stationary; (ii) the number $N$ of time series is fixed and
usually small. In this paper we relax both assumptions: the
$\{e_{i t}\}_{t = 1}^T$ can be non-stationary and $N$ can be
arbitrarily large. We describe the dependence of the $\{e_{i
t}\}_{t = 1}^T$ in terms of the physical dependence model of Wu
(2005), which represents errors as being generated by series of
i.i.d. innovations. The data-generating mechanism may be nonlinear
with respect to the innovation process and may vary smoothly over
time. This non-stationary dependence is realistic in practice and
it generalizes the parametric or stationarity assumptions of the
literature. Second, we devise a method of independent interest to
estimate consistently the long-run variance function of a locally
stationary time series. Third, we exploit a strong invariance
principle to build a simulation-based method that approximates the
finite sample distribution of the test statistic. The resulting
approximation is more accurate than the limiting normal
distribution and its implementation is faster than bootstrap
alternatives. Fourth, we apply the test to an iterative clustering
algorithm that groups time series according to the parallelism of
their trends. The algorithm has the nice feature that it does not
require to pre-specify in how many clusters the data will be
grouped. Time series that are very different from all others may
form a group of their own. For this reason, the algorithm provides
valuable insights in the data that complement standard approaches
like $k$-means clustering. Another attractive by-product of the
clustering algorithm is that it readily provides significance
levels for all clusters found.

The rest of the paper is organized as follows. Section
\ref{sec:teststat} presents a test statistic based on the
$L^2$-distances between the estimators of the individual trends
$(\mu_i-c_i)$ and the estimator of the global trend $\mu$ in
\eqref{eq:Null}. The test statistic estimates  a parallelism
index. Its asymptotic properties are discussed in Section
\ref{sec:asymp} for both fixed $N$ and $N\to\infty$. A central
limit theorem is derived under \eqref{eq:Null} and the test is
shown to be consistent against local alternatives. Section
\ref{sec:imple} deals with the test implementation and provides
methods for bandwidth selection and long-run variance estimation,
as well as a simulation-based method to approximate the
finite-sample distribution of the test statistic. Simulations are
carried out in Section \ref{sec:simu} to assess the empirical
significance level and statistical power of the test procedure.
The clustering algorithm is described  in Section
\ref{sec:clustering} and illustrated in Section \ref{sec:moto}
with the Motorola data. Proofs of the main results are deferred to
the Appendix.

\section{Test statistic}\label{sec:teststat}

To ensure model identifiability under the null hypothesis $H_0$ in
(\ref{eq:Null}), we assume that
\begin{eqnarray}\label{eq:identi}
\sum_{i=1}^N c_i = 0.
\end{eqnarray}
A natural way to test $H_0$ is to compare the curves $\hat \mu_i$
estimated under the general model (\ref{eq:Xn}) to the curves
$\hat c_i + \hat \mu$ estimated under $H_0$. To estimate the
common trend $\mu$ under $H_0$, we can use the averaged process
$\bar X_{\cdot t} = \sum_{i=1}^N X_{i t} / N$ for  $t=1, \ldots,
T$:
\begin{eqnarray}\label{eq:Avemu}
\bar X_{\cdot t} = \mu(t/T) + \bar e_{\cdot t}.
\end{eqnarray}
Similarly, define $\bar X_{i \cdot} = \sum_{t=1}^T X_{i t}
/ T$, $\bar X_{\cdot \cdot}$, $\bar e_{\cdot t}$ and $\bar e_{i
\cdot}$. In this paper we adopt the popular
local linear smoothing procedure (Fan and Gijbels, 1996) to
estimate the trends. Let $K$ be a Lipschitz continuous, bounded,
symmetric kernel function with support $[-1, 1]$ and satisfies $\int_{-1}^1 K(u) du = 1$; let $b > 0$ be
the bandwidth. Then the local linear estimator of $\mu$ is
\begin{eqnarray}\label{eq:hatmu}
\hat \mu(u) = \sum_{t=1}^T w_{b}(t,u)\bar X_{\cdot t} ,
\quad 0 \le u \le 1,
\end{eqnarray}
with the weights $w_b$ defined by
\begin{eqnarray}\label{eq:weights}
w_{b}(t,u) = K( (u-t/T)/ b)
 { {S_{b,2}(u) - (u-t/T) S_{b,1}(u)}
 \over { S_{b,2}(u) S_{b,0}(u) - S_{b,1}^2(u)}},
\end{eqnarray}
where
\begin{eqnarray}\label{eq:Sbj}
 S_{b,j}(u) = \sum_{t=1}^T (u-t/T)^j K( (u-t/T)/ b),
 \, \, u \in [0, 1].
\end{eqnarray}
Let $(\hat \beta_0, \hat \beta_1)$ be the minimizer of the
weighted sum
\begin{eqnarray*}
 \sum_{t=1}^T (\bar X_{\cdot t} - \beta_0 - \beta_1(u-t/T))^2
 K( (u-t/T)/ b).
\end{eqnarray*}
Then $\hat \mu(u) = \hat \beta_0$, and Fan and Gijbels (1996)
argued that this local linear estimate has a nice boundary
behavior. For simplicity of the procedure, it is advantageous to
estimate $\mu_i$ with the same bandwidth used for $\mu$. This also
simplifies mathematical derivations (see Section
\ref{sec:bandselec}). In this case, the local linear estimate for
$\mu_i$ is
\begin{eqnarray}\label{eq:mui}
\hat \mu_i(u) = \sum_{t=1}^T w_{b}(t,u) X_{i t}.
\end{eqnarray}
The intercepts $c_i$ are estimated by
\begin{eqnarray}\label{eq:hatci}
\hat c_i = {1\over T} \sum_{t=1}^T
 [\hat \mu_i(t/T) - \hat \mu(t/T)].
\end{eqnarray}
Since the same bandwidth $b$ is used in (\ref{eq:hatmu}) and
(\ref{eq:mui}), we have the interesting observation that $\hat
\mu(u) = N^{-1} \sum_{i=1}^N \hat \mu_i(u)$. Therefore, the $\hat
c_i$ naturally satisfy the constraint (\ref{eq:identi}).

There are many ways to measure the distance between the curves
$\hat c_i + \hat \mu(\cdot)$ and $\hat \mu_i(\cdot)$. In this paper we
adopt the $L^2$-distance
\begin{eqnarray}\label{eq:Delta2}
\widehat{\Delta}_{N,T} = \sum_{i=1}^N \int_0^1
 (\hat \mu_i(u) - \hat c_i - \hat \mu(u))^2 d u.
\end{eqnarray}
Clearly $\widehat{\Delta}_{N,T}$ is a natural estimate for the parallelism
index
\begin{equation}\label{Delta}
\Delta_N = \min_{\substack{\mu,c_1, \ldots, c_N \\
\sum_i c_i = 0}} \: \sum_{i=1}^N \int_0^1
 ( \mu_i(u) -  c_i - \mu(u))^2 d u,
\end{equation}
where the explicit solutions are $\mu(u) = \sum_{i=1}^N \mu_i(u) / N$ and
$c_i = \int_0^1 (\mu_i(u) - \mu(u)) d u$.

\section{Asymptotic theory}\label{sec:asymp}

Here we shall discuss limiting distribution and consistency of the
test. In our framework we allow both $N$ and $T$ to go to
infinity, and the error processes $\{e_{it}\}_{t=1}^T$ can be
non-stationary. To establish the asymptotic normality of
$\widehat{\Delta}_{N,T}$, we impose structural conditions on the
error processes $\{e_{it}\}_{t=1}^T$, following the ideas of Wu
(2005). More specifically, we assume that the $\{e_{it}\}_{t=1}^T,
\, i=1, \ldots, N,$ are  i.i.d.  as a process $\{e_t\}_{t=1}^T$ of
the form
\begin{eqnarray}\label{eq:et}
e_{t} = G(t/T; {\cal F}_{t}), 
\end{eqnarray}
where ${\cal F}_{t} =  (\ldots, \varepsilon_{t-1},\varepsilon_t)$,
$\{\varepsilon_j\}_{ j\in \Z} $ is an innovation process with
i.i.d. elements, and $G(\cdot;\cdot)$ is a measurable function.
Equation \eqref{eq:et} can be interpreted as an input/output
physical system where the $\{\varepsilon_j\}_{j=-\infty}^t$ are the
inputs and $e_{t} $ is the output. Assuming that $G(u; {\cal
F}_{t})$ has a finite $p$-th moment for some $p > 0$, define  the
physical dependence measure
\begin{equation}\label{eq:pdm}
\delta_p(t) = \sup_{0 \le u \le 1}
 \|G(u; {\cal F}_{t}) - G(u; {\cal F}'_{t}) \|_p,
\end{equation}
where $ {\cal F}'_t = (\ldots, \varepsilon_{-1}, \varepsilon'_{0},
\varepsilon_{1}, \ldots, \varepsilon_{t})$ and $\varepsilon'_{0}$
is a random variable such that $\varepsilon'_{0}$,
$\varepsilon_t$, $t \in \mathbb{Z}$, are i.i.d. The index
$\delta_p(t)$ quantifies the dependence of the output $e_t$ on the
inputs $\mathcal{F}_t$ by measuring the distance between
$G(\cdot;\mathcal{F}_t)$ and its coupled version $G(\cdot ;
\mathcal{F}'_t)$. Furthermore, assume that
$G(u; \mathcal{F}_t)$
is stochastically Lipschitz continuous (SLC), that is, there
exists a constant $C$ such that
\begin{eqnarray}\label{eq:Jun953}
\|G(u_1;\mathcal{F}_t) - G(u_2;\mathcal{F}_t)\|_p \le C |u_1 - u_2|
\end{eqnarray}
for all $u_1, u_2 \in[0,1]$, which we denote by $G \in SLC$. This
models the non-stationarity in which the underlying data
generating mechanism changes smoothly over time. Note that the
$\{e_{it}\}_{t=1}^T$ can be represented in the following manner:
let $\varepsilon_{i k}, \, i=1,\ldots,N, \, k \in \Z$, be i.i.d.
random variables; let ${\cal F}_{it} =  (\ldots, \varepsilon_{i,
t-1}, \varepsilon_{it})$, then
\begin{eqnarray}\label{eq:eit}
e_{i t} = G(t/T; {\cal F}_{i t}).
\end{eqnarray}
Assuming that $\E e_{k} = 0$ for all $k\in \Z$, let
\begin{equation}\label{th autocovariance}
\gamma_k(u) = \E [ G(u; {\cal F}_{k}) G(u; {\cal F}_0)],
 \quad 0 \le u \le 1.
\end{equation}
Define the long-run variance function
\begin{eqnarray}\label{eq:gt}
g(u) = \sum_{k \in \Z} \gamma_k(u)
\end{eqnarray}
and its squared integral
\begin{equation}\label{sigma_squared}
\sigma^2 = \int_0^1 g^2(u)du .
\end{equation}
Recall that the kernel function $K$ is Lipschitz continuous
on its support $[-1, 1]$. Let
\begin{equation*}
K^*(x) = \int_{-1}^{1-2 |x|} K(v) K(v+2 |x|) d v
\quad \textrm{and} \quad  K_2^* = \int_{-1}^1 ( K^*(v))^2 d v.
\end{equation*}

We have the following result.

\begin{theorem}
\label{th:Delta} Let $N=N(T)$ be such that either (i) $N \to
\infty$ as $T \to \infty$ or (ii) $N $ is fixed. Let $b = b(T)$ be
a bandwidth sequence such that $T b^{3/2} \to \infty$ and $b \to
0$. Further assume that $G \in SLC$ and that, for some $p > 4$,
the following short-range dependence condition holds:
\begin{eqnarray}\label{eq:srd}
\sum_{t = 0}^\infty \delta_p(t) < \infty.
\end{eqnarray}
Then under the null hypothesis $H_0$, we have
\begin{eqnarray}\label{eq:cltallN}
T b^{1/2} (N-1)^{-1/2}
 (\widehat{\Delta}_{N,T} - \E \widehat{\Delta}_{N,T})
 \mathop{\rightarrow}^{\mathcal{L}} N(0, \sigma^2 K_2^*).
\end{eqnarray}
\end{theorem}

It is worth observing that the limit distribution in
(\ref{eq:cltallN}) is the same whether (i) $N\to\infty$ or  (ii)
$N=O(1)$. However, the proofs for these two cases are different;
see Section \ref{sec:Thm1} in the Appendix. Here we provide
intuitions of the proofs. If $N \to \infty$, the estimates $\hat
c_i$ and $\hat \mu$ will both be close to their true values. Hence
the $\int_0^1 (\hat \mu_i(u) - \hat c_i - \hat \mu(u))^2 du, \,
i=1, \ldots, N,$ in (\ref{eq:Delta2}) can be approximated by the
$\int_0^1 (\hat \mu_i(u) - c_i - \mu(u))^2 du$, which are i.i.d.,
and the classical Lindeberg-Feller Central Limit Theorem (CLT)
applies.
In case (ii), the Lindeberg-Feller CLT is no longer applicable
since $N = N(T)$ is bounded; however, we can apply the
$m$-dependent and martingale approximations as in Liu and Wu
(2010) and still obtain (\ref{eq:cltallN}). Note that the factor
($N-1$) in (\ref{eq:cltallN}) is due to the fact that we average
the $N$ independent streams to get the function estimate $\hat
\mu$, thus losing one degree of freedom.

We now look into test consistency. Recall that $\widehat
{\Delta}_{N, T}$ serves as an estimate of the parallelism index
$\Delta_N$ defined in \eqref{Delta} under both $H_0$ in
\eqref{eq:Null} and alternatives. Our test rejects $H_0$ at level
$\alpha$ if $\widehat {\Delta}_{N,T}$ exceeds the $(1-\alpha)$
quantile of its distribution. (The precise implementation of the
test is provided in Section \ref{sec:imple}). The next theorem
asserts that this test is consistent against local alternatives
approaching \eqref{eq:Null} such that $N^{-1} (Tb + b^{-2})
\Delta_N \to \infty$, namely under the latter condition the power
goes to $1$.

\begin{theorem}
\label{thm:power} Assume conditions of Theorem \ref{th:Delta}.
Also assume that the $\mu_i,\, i=1,\ldots,N,$ in \eqref{eq:Xn}
have uniformly bounded second derivatives on $[0,1]$. Then the
parallelism test based on $\widehat{\Delta}_{N,T}$ has unit
asymptotic power if $N^{-1} (Tb + b^{-2}) \Delta_N \to \infty$.
\end{theorem}


\section{Test implementation}
\label{sec:imple}

We address here the implementation of Theorem \ref{th:Delta} for
hypothesis testing. In particular, we discuss the issues of
bandwidth selection and variance estimation, and we propose a
simulation-based procedure that improves upon the normal
approximation for the test statistic $\widehat{\Delta}_{N,T}$.

\subsection{Bandwidth selection}\label{sec:bandselec}


As seen in Section \ref{sec:teststat}, the same bandwidth $b$ is used in the test procedure to estimate both $\mu $ and the $\mu_i, \, i=1,\ldots,N$.
In addition to simplifying the test implementation and theoretical study, this choice automatically corrects biases under $H_0$ as noted by H\"ardle and Mammen (1993):
\begin{align}
\E [\hat \mu_{i}(v) - \hat \mu(v)] & = \sum_{t=1}^T w_b(t, v) \left[ c_i +\mu(t/T) \right] -
 \sum_{t=1}^T w_b(t, v) \mu(t/T) \nonumber \\
 & = c_i .
\end{align}


To select the bandwidth $b$, we propose a generalized
cross-validation (GCV) procedure that can adjust for the
dependence of the time series. The simulation study of Section
\ref{sec:simu} suggests that our test procedure is reasonably
robust to the choice of $b$ and the GCV method \eqref{GCV}
performs reasonably well. Since our test procedure aims at
reconstructing the mean function differences $\mu_i - \mu , \:
i=1, \ldots, N,$ and assess whether they are constant over time,
it is natural to base the GCV score on the $\mathbf{Y}_i =
\{X_{it} - \bar{X}_{\cdot t}\}_{t = 1}^T$ rather than on the
original time series $\mathbf{X}_i = \{X_{it}\}_{t=1}^T$. Let
$\boldsymbol{\Gamma} = (\gamma_{t, t'})_{1 \leq t,t' \leq T}$,
where $\gamma_{t, t'} = \mathbb{E} (e_{t} e_{t'})$, be the
covariance matrix of the error process and let $ \mathbf{H}(b)$ be
the $T\times T$ ``hat" matrix associated to the local linear
smoother with bandwidth $b$. Denoting by $\widehat{\mathbf{Y}}_i =
\mathbf{H}(b) \mathbf{Y}_i $ the estimator of $\mu_i - \mu$ at the
design points, we propose to choose $b$ by minimizing the GCV
score
\begin{equation}\label{GCV}
\mathrm{GCV}(b) =
 \sum_{i = 1}^N \frac{( \widehat{\mathbf{Y}}_i
  - \mathbf{Y}_i)^\top
 \boldsymbol{\Gamma}^{-1}( \widehat{\mathbf{Y}}_i - \mathbf{Y}_i)}
{ (1 - \mathrm{tr}(\mathbf{H}(b))/T)^2} \, .
\end{equation}
We now consider the estimation of the covariance matrix
$\boldsymbol{\Gamma} = (\gamma_{t, t'})_{1 \leq t, t' \leq T}$.
Due to the local stationarity of $e_{i t}$, we use the local
linear smoothing (Fan and Gijbels, 1996) technique and naturally
estimate $\gamma_{t, t+k}$, $0 \le k < T$, by
\begin{equation}\label{eqn:May111}
\hat \gamma_{t, t+k} = {1 \over N}
 \sum_{i=1}^N \sum_{v=1}^{T-k}
 \hat e_{iv} \hat e_{i,v+k} w_b\{v/(T-k),t/(T-k)\},
\end{equation}
where $w_b(t,u)$ are the local linear weights defined by
(\ref{eq:weights}) with $T$ therein replaced by $T-k$, and $\hat
e_{iv} = X_{iv} - \hat \mu_i(v/T)$, $i = 1,\ldots,N$, $v =
1,\ldots,T$, are the estimated residuals. Since $\gamma_{t, t'}$
is small if $|t-t'|$ is large, using the regularization method of
banding (Bickel and Levina, 2008), we estimate $\boldsymbol
\Gamma$ by $(\hat \gamma_{t, t'} I_{\{|t-t'| \leq T^{4/15}\}})_{1
\leq t, t' \leq T}$.

\subsection{Estimation of the long-run variance function}
\label{sec:longrun}

In order to apply Theorem \ref{th:Delta}, we
need to estimate the critical quantity $\sigma^2 $ in
\eqref{sigma_squared} which serves as the asymptotic variance (up
to a known scalar) of the test statistic (\ref{eq:Delta2}), or
more essentially we need to estimate the long-run variance
function $g$. For each $u \in [0,1]$, let
\begin{eqnarray}
\mathcal{N}_{\tau}(u) = \{t: |t/T - u| \leq \tau \},
\end{eqnarray}
where
$\tau =\tau(T)$ is a window size satisfying $\tau \to 0$
and $T \tau \to \infty$ as $T\to\infty$.
The points of $\mathcal{N}_{\tau}(u)$, suitably rescaled by $1/T$,
become increasingly dense in $[u-\tau,u+\tau]$  as $T\to\infty$.
By the local stationarity (\ref{eq:Jun953}),
the process $\{e_{i t}\}_{t \in \mathcal{N}_{\tau}(u)}$ can be
approximated by the stationary process $( G(u, {\cal F}_{i t})
)_{t \in \mathcal{N}_{\tau}(u)}$ in the sense that
\begin{eqnarray}
\sup_{0 \le u \le 1} \max_{t \in \mathcal{N}_{\tau}(u)}
 \| e_{i t} - G(u, {\cal F}_{i t}) \|_p = O(\tau).
\end{eqnarray}
Denote by $\hat \gamma_{i k}(u)$ the sample auto-covariance of
$\{e_{i t}\}_{t \in \mathcal{N}_{\tau}(u)}$ at lag $k$ and
average these quantities over $i$ to estimate the auto-covariance
\eqref{th autocovariance} by
\begin{eqnarray}\label{eq:}
\hat \gamma_k(u) = {1\over N} \sum_{i=1}^N \hat \gamma_{i k}(u).
\end{eqnarray}
Then $g(u)$ can be simply estimated by
\begin{eqnarray}\label{eq:ghat}
\hat g(u) = \sum_{k=- K_T}^{ K_T} \hat \gamma_k(u)
\end{eqnarray}
for some truncation parameter $K_T= \lfloor T \tau \varrho
\rfloor$ with bandwidth $\varrho \to 0$ and $T \tau \varrho \to
\infty$. Indeed, $\gamma_k(u)$ will be close to zero for large $k$
and for all $u\in [0,1]$ under the local stationarity condition
\eqref{eq:Jun953} and the short-range dependence assumption
(\ref{eq:srd}).
More precisely, we need  to specify the decay rate of the physical
dependence measure (\ref{eq:pdm}) to characterize the bias caused
by truncation. Also, the error processes $\{e_{it}\}, \, i=1,\ldots,
N,$ are not observable in practice and we recommend plugging the
residuals $\hat e_{it} = X_{i t} - \hat \mu_i(t/T)$ from
\eqref{eq:mui} into (\ref{eq:ghat}) to get an estimate $\tilde g$
of the long-run variance function. The following theorem provides
error bounds  for both $\hat g$ and $\tilde g$.

\begin{theorem}\label{thm:ghat}
Assume that $G \in SLC$, $g \in \mathcal{C}^2[0,1]$,
$\sum_{t=0}^\infty \delta_{4}(t)  < \infty$, and
$\sum_{t=T}^\infty \delta_{2}(t) = \mathcal{O}(T^{-\alpha})$ for
some $\alpha > 0$. Then
\begin{equation}\label{eqn:bndghat}
\sup_{u \in [0,1]} \| \hat g(u) - g(u) \|_{2}
 = \mathcal{O}\left (\sqrt{\varrho / N} + (T \tau \varrho)^{-\alpha}
 + (\tau\rho)^{\alpha/(1+\alpha)} + \tau^2 + \varrho\right).
\end{equation}
If in addition $\iota = (T \tau \varrho)^{1/2} (b^2 + T^{-1/2}
b^{-1/2}) \to 0$, we have
\begin{equation}\label{eqn:bndgtilde}
\sup_{u \in [0,1]} \| \tilde g(u) - g(u) \|_{2}
 = \mathcal{O}\left (\iota + \sqrt{\varrho / N}
 + (T \tau \varrho)^{-\alpha}
 + (\tau\rho)^{\alpha/(1+\alpha)} + \tau^2 + \varrho
 \right).
\end{equation}
\end{theorem}

The choice of banding parameters $\tau$ and $\varrho$ that
minimize the bound on the right hand side of (\ref{eqn:bndghat})
can depend on $N$, $T$ and $\alpha$ in a highly complicated
fashion. Nevertheless, when $\alpha \geq 2$ we have the following
dichotomy:
\begin{itemize}
\item If $N \geq T^{2\alpha / (3\alpha + 2)}$, the optimal bound
in (\ref{eqn:bndghat}) is $\mathcal{O}(T^{-2\alpha /(3\alpha +
2)})$ for $\tau \asymp T^{-\alpha /(3\alpha + 2)}$ and $\varrho
\asymp T^{-2\alpha /(3\alpha + 2)}$;

\item If $N \leq T^{2\alpha / (3\alpha + 2)}$ in which case $N$ is
not required to blow up, the optimal bound in (\ref{eqn:bndghat})
is  $\mathcal{O}((TN)^{-2\alpha /(5\alpha +
2)})$ for $\tau \asymp (TN)^{-\alpha /(5\alpha + 2)}$ and $\varrho
\asymp T^{-4\alpha /(5\alpha + 2)} N^{(\alpha + 2) /(5\alpha +
2)}$.
\end{itemize}
In particular when the errors satisfy the geometric moment
contraction condition, that is, $\delta_{2}(k)$ decays
geometrically quickly as in the case of an autoregressive process,
the optimal bound for (\ref{eqn:bndghat}) is $\mathcal{O}(T^{-2/3}
\log T)$ if $N/T^{2/3} \to \infty$ and $\mathcal{O}(T^{-2/5} \log
T)$ otherwise.

Note that the bound in (\ref{eqn:bndgtilde})
goes to zero at a slower rate than the one in (\ref{eqn:bndghat})
and reaches $\mathcal{O}(T^{-2/5}\log T)$ when the geometric
moment contraction condition is satisfied.

\subsection{Simulation-based approximation to the distribution of
the test statistic}\label{sim-basedapprox}

The normal convergence in (\ref{eq:cltallN}) can be quite slow. A
popular way to improve the convergence speed is via bootstrap; see
for example Hall and Hart (1990) and Vilar-Fernandez et al. (2007).
Here we propose an alternative simulation-based method, which is
easily implementable and has a better finite-sample performance.

Let $Z_{it}$, $i = 1,\ldots, N$, $t = 1, \ldots, T,$ be i.i.d.
standard normal random variables. If the long-run variance
function $g$ is known, let $X_{it}^\diamond = g(t/T) Z_{it}$ and
otherwise, use the estimate $\tilde g$ to define $X_{it}^\diamond
= \tilde g(t/T) Z_{it}$. Let $\widehat{\Delta}_{N,T}^\diamond$ be
the test statistic associated to the $X_{it}^\diamond$, assuming
that $c_i \equiv 0$. By Theorem \ref{th:Delta}, $\widehat
{\Delta}_{N, T}$ and $\widehat {\Delta}_{N, T}^\diamond$ have the
same asymptotic distribution under the assumptions of Theorems
\ref{th:Delta} and \ref{thm:power}. Hence, the distribution of
$\widehat {\Delta}_{N, T}$ can be assessed by simulating $\widehat
{\Delta}_{N, T}^\diamond$. Specifically, one can generate many
realizations of $(X_{it}^\diamond)_{t=1}^T, i=1, \ldots N,$ and
compute the corresponding $\widehat{\Delta}_{N,T}^\diamond$ from
which one can obtain the estimated $(1-\alpha)$-th quantile $\hat
q_{1-\alpha}$. Based on this, one can reject at level $\alpha$ the
null hypothesis if $\widehat{\Delta}_{N,T} > \hat q_{1-\alpha}$,
and accept otherwise. The validity of this method is guaranteed by
the invariance principle (see Wu and Zhou (2011)) which asserts
that partial sums of dependent random vectors can be approximated
by Gaussian processes.


\section{Simulation study}\label{sec:simu}
\subsection{Acceptance Probabilities}
\label{subsec:acceptprob} In this section we present a simulation
study to assess the performance of our test procedure. Consider
the model
\begin{eqnarray}\label{mod:sim}
X_{i t} = c_i + \mu(t/T) + e_{i t},
\end{eqnarray}
with $c_i = 0$ and $\mu(u) = 2 \sin (2\pi u)$. Note that under
\eqref{mod:sim}, the test procedure is independent of the $c_i$.
The error process $\{e_{it}\}$  is generated by $e_{i t} =
\zeta_{i,t}(t/T)$, where for all $i \in \mathbb{Z}$ and $u \in
[0,1]$, the process $(\zeta_{i,t}(u))_{t \in \mathbb{Z}}$ follows
the recursion
\begin{eqnarray}\label{eqn:simu}
\zeta_{i,t}(u) = \rho(u) \zeta_{i,t-1}(u)
 + \sigma \varepsilon_{i,t},
\end{eqnarray}
with the $\varepsilon_{i,t}$ being i.i.d. random variables
satisfying $\mathbb{P} (\varepsilon_{i,t} = -1) = \mathbb{P}
(\varepsilon_{i,t} = 1) = 1/2$. Thus, $\{e_{it}\}_{t \in
\mathbb{Z}}$ for $i = 1, \ldots, N$ are i.i.d. AR(1) processes
with time-varying coefficients. Let $\rho(u) = 0.2 - 0.3u$ and
$\sigma = 1$. Easy calculations show that $\mathbb{E}
(\zeta_{i,t}(u)) = 0$, $\mathrm{Var}(\zeta_{i,t}(u)) = \sigma^2 /
(1-\rho(u)^2)$ and the long-run variance function $g(u) = \sigma^2
/ (1-\rho(u))^2$.

In our simulation the Epanechnikov kernel $K(v) = 3 \max(0,
1-v^2)/4$ is used. We simulate 10,000 realizations of
(\ref{eqn:simu}) and, for each realization, 10,000 simulations of
$\widehat{\Delta}_{N,T}^\diamond$ are performed as in Section
\ref{sim-basedapprox}. We are interested in the proportion of
realizations for which the null hypothesis is correctly accepted.
Acceptance probabilities are presented in Table
\ref{tab:coverageprob} for different choices of $T$, $N$ and $b$.
This suggests that the acceptance probabilities are reasonably
close to the 95\% nominal levels and become more robust to the
size of bandwidth as the sample size gets bigger.

\begin{table}[ht]
\centering \caption{Acceptance probabilities at 95\%
nominal levels with different $T$, $N$ and $b$.}\label{tab:coverageprob}
\begin{tabular}{l|lll|lll|lll}
\hline \hline \multicolumn{1}{c}{} & \multicolumn{3}{c}{$T = 100$}
 & \multicolumn{3}{|c}{$T = 300$} & \multicolumn{3}{|c}{$T = 500$} \\
$b \backslash N$ & 50 & 100 & 150 & 50 & 100 & 150 & 50 & 100 &150 \\
\hline
.1 & .977 & .979 & .979 & .955 & .963 & .963 & .955 & .959 & .959 \\
.2 & .969 & .970 & .973 & .947 & .960 & .960 & .959 & .955 & .956 \\
.3 & .962 & .964 & .969 & .949 & .957 & .958 & .955 & .952 & .955 \\
.4 & .958 & .966 & .962 & .954 & .951 & .957 & .958 & .954 & .956 \\
.5 & .961 & .959 & .963 & .954 & .956 & .959 & .948 & .958 & .948 \\
.6 & .955 & .959 & .959 & .952 & .953 & .949 & .950 & .945 & .958 \\
.7 & .952 & .964 & .962 & .949 & .958 & .951 & .948 & .954 & .953 \\
.8 & .958 & .963 & .962 & .953 & .951 & .953 & .953 & .953 & .947 \\
.9 & .957 & .959 & .963 & .955 & .956 & .950 & .950 & .950 & .953 \\
\hline \hline
\end{tabular}
\end{table}

\subsection{Statistical power}\label{subsec:asymppower}

\begin{figure}[b]
\begin{center}
\includegraphics[width= .8\textwidth]{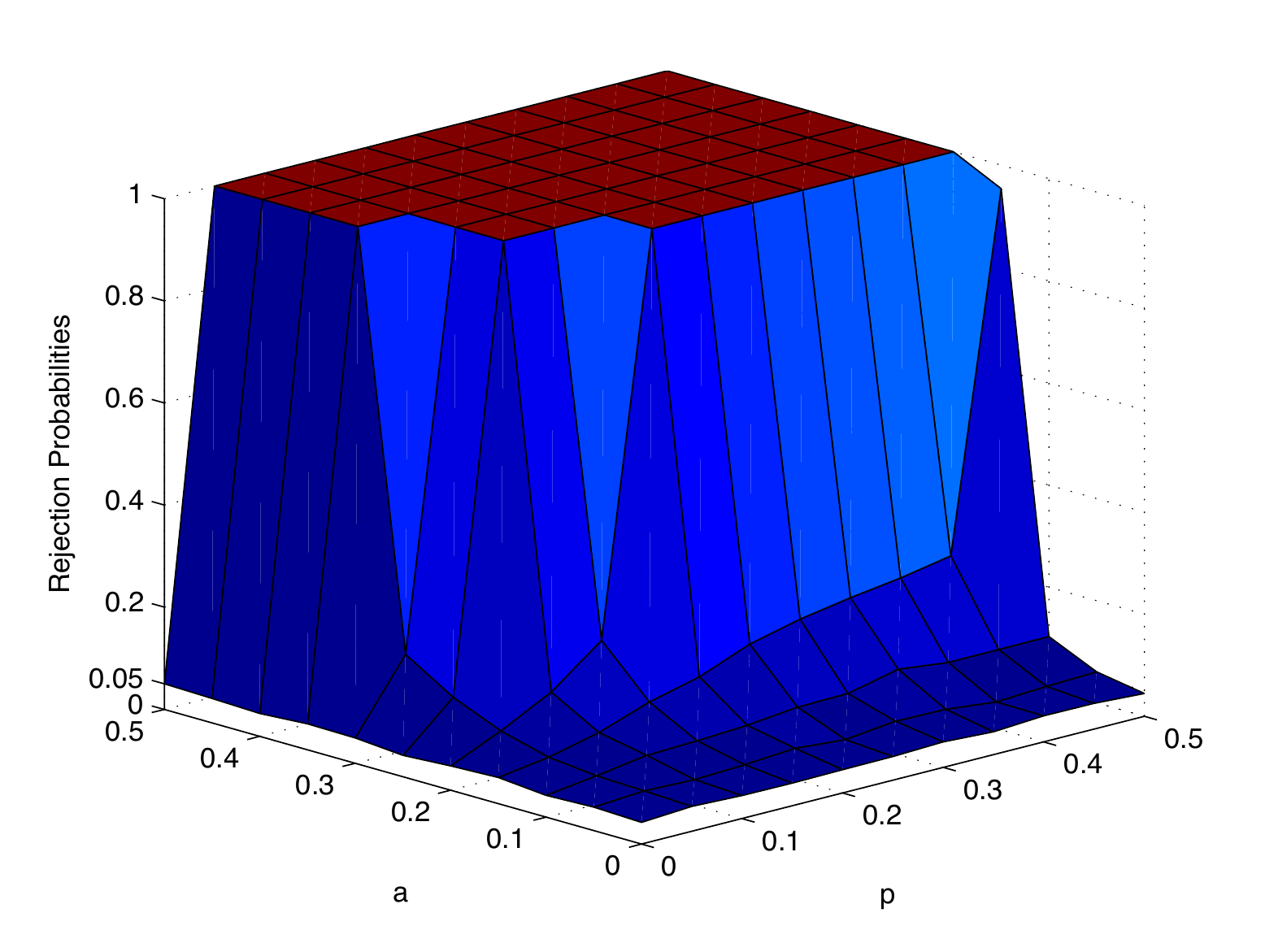}
\caption{Rejection probabilities at 5\% nominal levels with different choices of deviation proportion $p$ and distortion magnitude $a$.} \label{fig:simupower}
\end{center}
\end{figure}

In the setting of Section \ref{subsec:acceptprob} with $T = 300$ and $N = 100$, we study the statistical power of our testing procedure. For a certain proportion (say $p$) of the $N$ time series $\{X_{it}\}_{t=1}^T$ for $i \in \{1,\ldots,N\}$, we add a distortion $a \mu_d(t/T)$ in addition to (\ref{mod:sim}), where $\mu_d(u) = 2 \cos(2 \pi u)$ and $a$ denotes the corresponding magnitude. We investigate on the rejection probabilities at 5\% nominal levels with different choices of $p$ and $a$ and the results are summarized in Figure \ref{fig:simupower}. It can be easily seen that the power goes to one very quickly as the magnitude of distortion $a$ gets large and the proportion of different trends $p$ approaches 0.5.

\section{Application of the test to clustering}\label{sec:clustering}

The test procedure developed in the previous sections can be
applied to cluster collections of time series based on their
similarity in terms of parallelism. In the sequel we identify the
time series in \eqref{eq:Xn} with their indexes. To build our
iterative clustering algorithm, we start by finding the largest
cluster $G_1$ in $U^{(0)} = \{ 1,\ldots,N\}$ for which the
parallelism assumption $H_0$ is retained at  level $\alpha$. The
cluster $G_1$ is obtained by progressively removing from the
analysis the sample units that contribute most to the test
statistic \eqref{eq:Delta2}. Specifically, $H_0$ is first tested
on $U^{(0)}$, then on a subset $U^{(1)} \subset U^{(0)}$ if
rejected on  $U^{(0)}$, and so on so forth until $H_0$ is accepted
or $U^{(k)}$ is reduced to a single element for some $k$, in which
case the algorithm ends without any cluster being found. At the
second iteration, the procedure is repeated with the remaining
time series (set $U^{(0)}:= \{ 1,\ldots,N\} \setminus G_1$) and so
on so forth until either all time series are clustered (i.e.
$\{1,\ldots,N\} = G_1 \cup \cdots \cup G_L$ for some $L$) or no
more clusters of size $>1$ can be formed.

We now give a precise description of the algorithm. For the test
implementation, the user must provide a significance level
$\alpha$, bandwidth $b$ (cf. Sections \ref{sec:teststat} and
\ref{sec:bandselec}), and parameters $(\tau,\varrho)$ (cf. Section
\ref{sec:longrun}). The user must also specify the number $n$ of
sample units to remove at each step of the cluster search. As long
as $n$ is small (say $n=1$ for small $N$ and, say $n \le 5N / 100
= N/20$ for moderate to large $N$), this tuning parameter does not
affect the outcome of the clustering algorithm; it however
influences the computational time. Note that when moving from
working index set  $U^{(k)} $ to $U^{(k+1)} $ during the cluster
search, the algorithm removes at least one and at most $(N_k-2)$
sample units  from $U^{(k)} $, where $N_k$ is the size of $U^{(k)}
$, so that there remains at least two units to compare at the next
step. As a result the effective number of removed units is $n^\ast
= \max (1, \min(n,N_k-2 ))$. Also, if $H_0$ is rejected on
$U^{(k)} $ and accepted on $U^{(k+1)} $, this may mean that too
many units ($n^\ast$ of them) have been removed from $U^{(k)} $
and that $H_0$ can be retained on an intermediate set $U^{(k+1)}
\subseteq U' \subset U^{(k)} $. In this case a flag $F$ is
activated and the algorithm starts a dichotomic search, returning
to the previous working index set  $U^{(k)} $ and attempting to
remove less units at each subsequent step (i.e. roughly $n/2$,
then $n/4$, etc.). The following notations are needed for the
formal statement of the algorithm:
\begin{itemize}
\item[$\ast$] $k$: step counter; $l$: group counter, $F$: flag.
\item[$\ast$] $\bar{X}^{(k)}_{\cdot t} = N_k^{-1} \sum_{i\in
U^{(k)}} X_{it}$, $\bar{X}_{\cdot \cdot}^{(k)} = T^{-1}
\sum_{t=1}^T \bar{X}^{(k)}_{\cdot t} $, $\hat \mu^{(k)}(u) =
\sum_{t=1}^T w_{b}(t,u) \bar{X}_{\cdot t}^{(k)}$, and
$\hat{c}_i^{(k)} = T^{-1} \sum_{i\in U^{(k)}} \big( \hat
\mu_i(t/T) - \hat \mu^{(k)}(t/T) \big) $. Recall that $N_k$ is the
size of  $U^{(k)} $.
\end{itemize}

The algorithm works as follows:
\bigskip

\noindent \textbf{Initialization.}

\begin{enumerate}

\item Set $U^{(0)} := \{ 1, \ldots,N \}$, $k:=0$, $l:=1$, and $F:=0$.

\item Initialize the parameters $\alpha, b,\tau, \varrho, $ and
$n$, with $n<N_0$.

\item Perform the parallelism test on $\{X_{it}\}_{t=1}^T$ for $
i\in U^{(0)}$ and compute the $p$-value.
\begin{enumerate}
 \item Case $p > \alpha$.
 \begin{itemize}
\item  Compute the $\Delta_i := \int_0^1 \big\{ \hat{\mu}_i(u)
- \hat{c}_i^{(0)} - \hat \mu^{(0)}(u)\big\}^2du$ for $i \in U^{(0)}$
 and sort them as $\Delta_{\sigma(1)} \le \cdots \le \Delta_{\sigma(N_0)}$.
 \item Write $n^\ast := \max(1,\min(n,N_0-2))$ and
set \[ U^{(1)}:=U^{(0)} \setminus \{ \sigma(1),\ldots,
\sigma(n^\ast) \} \quad \textrm{and} \quad k:=1. \]
\end{itemize}

\item Case $p \le  \alpha$.
\begin{itemize}
\item   Set $G_1 := U^{(0)}$ and \underline{stop the algorithm}.
\end{itemize}
\end{enumerate}
\end{enumerate}

\medskip
\noindent \textbf{Determination of the cluster $G_l $.}
\begin{enumerate}
\setcounter{enumi}{3}

\item If $N_k=1 $, then \underline{stop the algorithm}.
\item Perform the parallelism test on the $\{X_{it}\}_{t=1}^T $
 for $ i\in U^{(k)}$ and compute the $p$-value.
\begin{enumerate}
 \item Case $p > \alpha$.
 \begin{itemize}
\item Compute the $\Delta_i := \int_0^1 \big\{ \hat{\mu}_i(u)
  - \hat{c}_i^{(k)} - \hat \mu^{(k)}(u)\big\}^2du$ for $i \in U^{(k)}$
 and sort them as $\Delta_{\sigma(1)} \le \cdots \le \Delta_{\sigma(N_k)}$.
\item If $F=1$, set $n:=\max(1,\lfloor n/2 \rfloor  )$.
\item Write $n^\ast := \max(1,\min(n,N_k-2))$ and set
\[ U^{(k+1)}:=U^{(k)} \setminus \{ \sigma(1),\ldots, \sigma(n^\ast) \}
  \quad \textrm{and} \quad k:=k+1.\]
\item Return to step 4.
\end{itemize}

 \item Case $p\le \alpha$.
\begin{itemize}
\item Set $F:=1$.
\end{itemize}
\begin{enumerate}
\item[(i)] Case $n=1$.
\begin{itemize}
\item  Set $ G_l : = U^{(k)} . $
\item  If $\{ 1,\ldots,N \} = (G_1 \cup \cdots \cup G_l)$,
 then \underline{stop the algorithm}. \\
 Else return to step 1 and set
 \[ U^{(0)}:= \{ 1,\ldots,N \} \setminus (G_1 \cup \cdots \cup G_l),
 \quad k:=0, \quad \textrm{and} \quad l:=l+1 . \]
\end{itemize}

\item[(ii)] Case $n>1$.
\begin{itemize}
\item Set \[ n:= \max (1, \lfloor n/2 \rfloor )
 \quad \textrm{and} \quad U^{(k)}:=U^{(k-1)}
 \setminus \{ \sigma(1),\ldots, \sigma(n^\ast) \},\]
where $n^\ast := \max(1,\min(n,N_{k-1}-2))$.
\item Return to step 4.
\end{itemize}

\end{enumerate}

\end{enumerate}
\end{enumerate}

The R implementation of the algorithm can be obtained from the
authors upon request.


\section{Analysis of Motorola data}
\label{sec:moto}

\begin{figure}[b]
\begin{center}
\includegraphics[width= \textwidth]{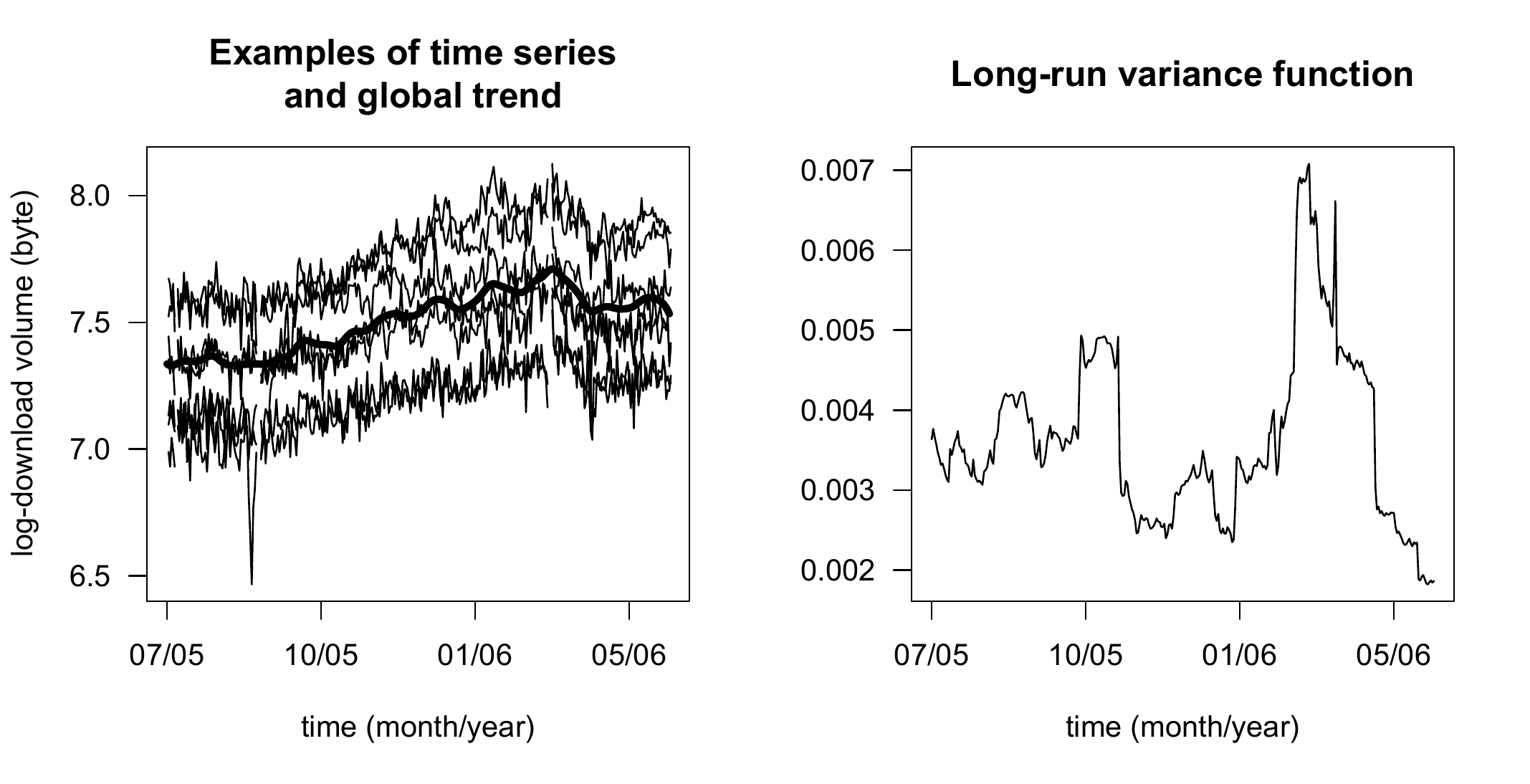}
\caption{Left panel: examples of download volume time series
(daily-totaled and log-transformed) and estimated global trend in
thick line. Right panel: Estimated long-run variance function. }
\label{fig:tsplot}
\end{center}
\end{figure}

To illustrate our parallelism test and clustering procedure, we
consider a data set of hourly volumes of downloads from cell
phones (in byte) in 129 U.S. area codes (24 area codes are in Center
America, 87 in Eastern America, 1 in Hawaii, and 24 in Pacific
America). Rather than studying the original data, we look into
their daily sums so as to remove daily periodicity (which produces
long-range dependence). Since the area codes have different
numbers of phone users, we also apply a logarithmic transform
(base 10) to the data to adjust for the scale effect. Thus,
multiplicative differences in the time series become additive,
which makes  it relevant to test for parallelism in the trends of
area codes. Examples of time series as well as the estimated
global trend function $\mu = N^{-1}\sum_{i=1}^N \mu_i$ and
long-run variance function $g$ are displayed in Figure
\ref{fig:tsplot}.

\begin{figure}[b]
\begin{center}
\includegraphics[scale=.65]{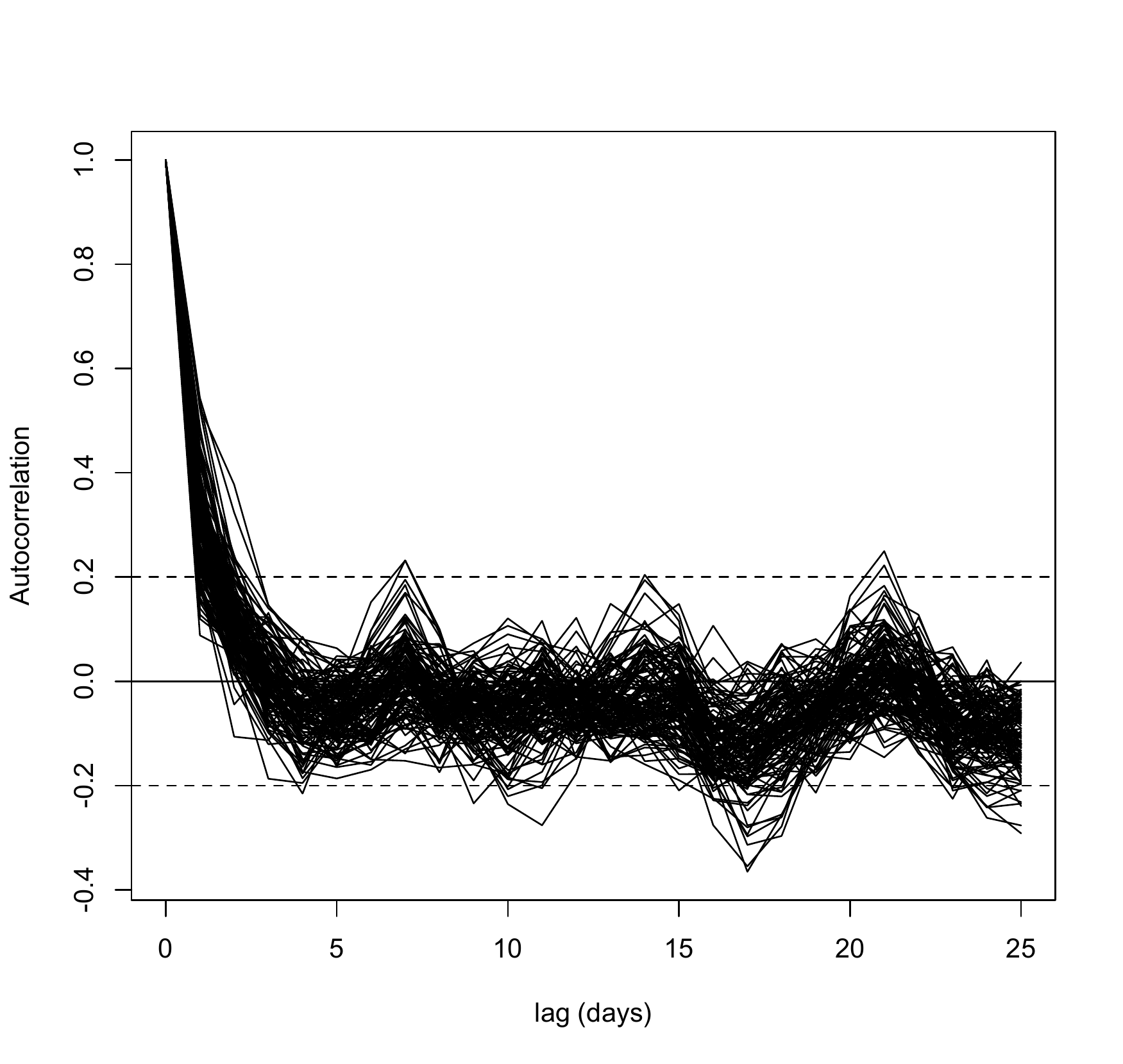}
\caption{Autocorrelation functions of the residual time series after a local linear fit with bandwidth $b=10$ days.}
\label{fig:acf}
\end{center}
\end{figure}

Prior to statistical analysis, the validity conditions of our
theoretical results have been verified on the data set. In
particular, the rapid decrease observed in the autocorrelation
functions of the detrended time series (see Figure \ref{fig:acf})
indicates that the short-range dependence assumption
\eqref{eq:srd} is very plausible. Also global similarity in the
autocovariance functions across the time series make the
assumption of identically distributed error processes look reasonable.
Finally, the $129\times 129$ cross-correlation matrix of the residual time
series has nearly zero entries outside its diagonal,
which suggests that the  time series are independent.

We now describe the implementation of the test and clustering
algorithm on the data set. The significance level of the test is
set to $5\%$. The local linear estimation of the trends is based
on a bandwidth $b$ and a truncated standard Gaussian density. The
inverse covariance matrix $\boldsymbol{\Gamma}^{-1}$ is estimated
as in Section \ref{sec:bandselec}. Specifically, after a pilot
trend estimation using a bandwidth $b=5$ days, the ``banding the
inverse covariance matrix" technique (e.g. Bickel and Levina
(2008)) has been applied to the sample covariance matrix of
residuals with a banding parameter $k=6$ days. The final bandwidth
$b$ is obtained by minimizing the GCV score \eqref{GCV}. The
long-run variance function $g$ is estimated as in Section
\ref{sec:longrun} based on the residuals of a local linear
smoothing with bandwidth $b=10$. (A larger $b$ is used to estimate
the long-run variance function than for the trend estimation so as
to make the estimate less sensitive to extreme observations.) The
parameters $\tau=0.04$ and $\varrho=0.31$ are chosen so that the
estimate of $g(u),u\in [0,1]$ utilizes about 2 weeks of data
before and after a time point $u$ and the autocovariances are
truncated at lag $K_T=4$. These parameter values are based on the
visual inspection of the autocovariance plots. Finally the number
$n$ of units to remove at each step of the cluster search (see
Section \ref{sec:clustering}) is set to 3, a good compromise
between search accuracy and computational speed of the algorithm.

\begin{figure}[htbp]
\begin{center}
\includegraphics[scale=.9]{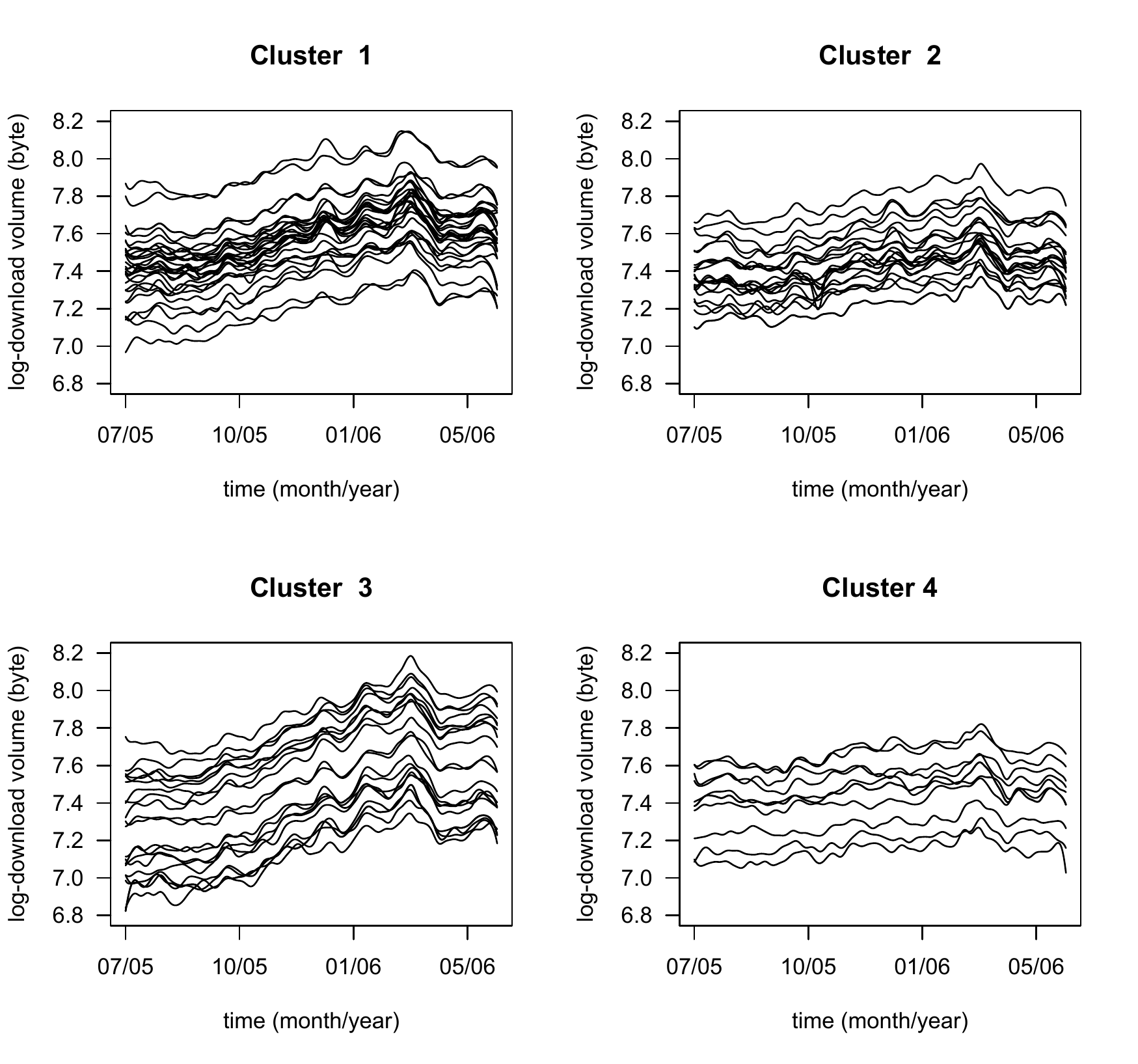}
\caption{Trends in the clusters of size $s\ge 10$. After a
log-transform of the data, the trends are obtained by smoothing
the time series with a bandwidth $b=4$ days. }
\label{clusters.pic}
\end{center}
\end{figure}

The results of the analysis are presented in Tables \ref{max
number of area codes retained} and 3. Note that performing the
parallelism test on the entire data set resulted in a $p$-value
$<10^{-16}$. Shifting our focus to clustering the time series, we
observe that the four largest clusters found contain respectively
29, 21, 20, and 10 area codes. This alone represents a sizable
proportion (62\%) of  the 129 area codes under study. These
clusters are displayed in Figure \ref{clusters.pic}, where their
homogeneity  can be observed. The examination of Table 3 reveals
that there is no obvious spatial pattern in the clusters. It also
shows that there is no systematic relation between the size of a
cluster and its associated $p$-value. Overall, our statistical
analysis shows that most area codes under study can be classified
in a small number of profiles, or clusters, according to the
parallelism patterns in their phone download activity. These
strong similarities across area codes would deserve to be
investigated in more detail as potentially, they could be
exploited by phone companies to e.g. better target their marketing
strategies or improve the bandwidth allocation.

\begin{table}[htdp]
\begin{center}
\begin{tabular}{|c|c| c|c|c|c|c|c|c|c|}
\hline
Cluster size & 29 &21& 20& 10&  7&  5&  3&  2&  1 \\
\hline
\# clusters & 1 &1 & 1 & 1 & 1 & 1 & 2 & 7 & 17\\
\hline
cum. prop. of $N$ & 22\% & 39\% & 54\%& 62\% & 67\% & 71\% & 76\% & 87\% & 100\%    \\
\hline
\end{tabular}
\end{center}
\caption{Summary of the clusters. The clusters are maximal sets of
area codes for which the parallelism assumption is retained at the
significance level 5\%.} \label{max number of area codes retained}
\end{table}%

\begin{table}[htdp]\label{clusters: spatial location}
{\footnotesize
\begin{center}
\textbf{Cluster 1: size 29, p-value 0.095}
\vspace*{2mm}

\begin{tabular}{|c c| c c | c c | c c| c c | c c | }
\hline
code & state & code & state &code & state &code & state &code & state& code & state \\
\hline
203&CT&  321&FL&  517&MI&  617&MA& 810&MI& 909&CA \\
219&IN& 323&CA&  540&VA&  619&CA&  813&FL&941&FL \\
239&FL& 484&PA&  562&CA&  646&NY& 815&IL&951&CA \\
301&MD&508&MA&  603&NH& 661&CA& 856&NJ& 978&MA \\
302&DE&513&OH&616&MI&703&VA& 859&KY& & \\
\hline
\end{tabular}

\vspace*{7mm}
\textbf{Cluster 2: size 21, p-value 0.064 }
\vspace*{2mm}

\begin{tabular}{|c c| c c | c c | c c| c c | c c | }
\hline
code & state & code & state &code & state &code & state &code & state& code & state \\
\hline
209&CA&586&MI&630&IL&732&NJ&781&MA& 908&NJ \\
240&MD&609&NJ&631&NY&734&MI&786&FL &&\\
510&CA&610&PA&708&IL&772&FL&818&CA &&\\
561&FL&626&CA&714&CA&774&MA &845&NY && \\
\hline
\end{tabular}

\vspace*{7mm}
\textbf{Cluster 3: size 20, p-value 0.151 }
\vspace*{2mm}

\begin{tabular}{|c c| c c | c c | c c| c c | }
\hline
code & state & code & state &code & state &code & state &code & state \\
\hline
231&MI& 404&GA& 512&TX& 803&SC&904&FL \\
269&MI& 407&FL& 704&NC&816&MO&919&NC\\
352&FL& 412&PA& 740&OH&863&FL&937&OH\\
386&FL& 419&OH& 773&IL&864&SC&989&MI\\
\hline
\end{tabular}

\vspace*{7mm}
\textbf{Cluster 4: size 10, p-value 0.071 }
\vspace*{2mm}

\begin{tabular}{|c c| c c | c c | c c| c c | }
\hline
code & state & code & state &code & state &code & state &code & state \\
\hline
  248&MI&570&PA&805&CA&  847&IL  &916&CA \\
516&NY&571&VA&808&HI&914&NY &917&NY \\
\hline
\end{tabular}
\vspace*{3mm}
\caption{Spatial locations of clusters. Only clusters of size $s\ge  10$ are displayed.}
\end{center}}
\end{table}


\section{Conclusion}

In this paper we have presented a test methodology for assessing
the parallelism between trends of multiple time series. The
physical dependence structure considered here allows to flexibly
model nonstationary time series without having to specify some
generating mechanism or autocovariance function. A method for
estimating the long-run variance function of locally stationary
processes and a simulation-based device to approximate the
distribution of statistics based on smoothed time series have been
developed as by-products of the test methodology. Both these tools
have shown good numerical performances in our simulations. They
are of independent interest and could be used with profit in other
statistical problems. A key assumption used to derive the theory
of  this paper is that the observed time series are independent
from one another. A very interesting extension would be to allow
for some form of dependence, e.g. to handle  spatio-temporal data.

The paper also proposes an innovative method to cluster time
series according to their parallelism properties. This method has
at least two attractive features: first, it does not require to
prespecify the number of clusters to be found, which guarantees
the homogeneity of the clusters and allows atypical time series to
be set apart; second, it readily provides significance levels for
each cluster, thereby giving a quantitative sense of how strong
the parallelism assumption holds. The implementation of this
clustering method has given meaningful results with the Motorola
time series. The algorithm is computationally fast as the
individual trend functions and long-variance functions need being
estimated only once, while the most computationally intensive step
(Gaussian process simulation to approximate the distribution of
the test statistic) is still manageable. The ideas harnessed in
this algorithm (greedy search, clustering based on individual
contribution of sample units to test statistic) can be used to
cluster time series according to other similarity measures than
parallelism. An interesting direction of future research would be
to compare the results of this type of clustering to the more
conventional $k$-means or hierarchical approaches.

\section{Appendix}
\label{sec:appendix} In the proofs we use $C$ to denote a constant
whose value may vary from place to place. It does not depend on
$N$ and $T$.

\subsection{Proof of Theorem \ref{th:Delta}}
\label{sec:Thm1} The techniques for handling Case (i) with large
$N$ and Case (ii) with fixed $N$ are different. For the former we
apply the traditional Lindeberg-Feller CLT, while for the latter,
we apply the $m$-dependence and martingale approximation
techniques. For details see Sections \ref{sec:bigN} and
\ref{sec:smallN}, respectively.

We start by showing that under $H_0$,
the test statistic $\widehat{\Delta}_{N,T}$ does not depend upon $\mu(\cdot)$ nor the $c_i$.
To see this, introduce the weight averages
\begin{equation*}
\bar w_b(t) = T^{-1} \sum_{j=1}^T w_b(t, j/T)
\end{equation*}
With \eqref{eq:hatmu}, \eqref{eq:mui}, and (\ref{eq:hatci}), we
easily see that
\begin{align*}
 \hat \mu_i(u) - \hat \mu(u) - \hat c_i
 & = \sum_{t=1}^T \left(w_{b}(t,u) - \bar w_b(t)\right)
  \left(c_i + e_{i t} - \bar e_{\cdot t}\right)
\\
& = \sum_{t=1}^T  \left\{ w_{b}(t,u) - \bar w_b(t)\right\}
 \left(e_{i t} - \bar e_{\cdot t}\right).
\end{align*}
The last equality stems from the fact that $ \sum_{t=1}^T
\left[w_b(t, u) - \bar w_b(t)\right] = 1-1=0$ by the well known
property that the weight functions $w_b(t,\cdot),t=1,\ldots,T,$
 of the local linear smoother sum up to one.

\subsubsection{Case (i): $N \to \infty$}
\label{sec:bigN} We shall prove the asymptotically
equivalent form of (\ref{eq:cltallN})
\begin{eqnarray}\label{eq:cltD2}
T b^{1/2} N^{-1/2} (\widehat{\Delta}_{N,T}- \E \widehat{\Delta}_{N,T})
 \mathop{\rightarrow}^{\mathcal{L}} N(0, \sigma^2 K_2^*).
\end{eqnarray}
To this end, we use the decomposition
\begin{equation}\label{eq:May12}
\widehat{\Delta}_{N,T}- \E \widehat{\Delta}_{N,T} = \sum_{i=1}^N (A_i - \E A_i )
 - \left( R_N - \E R_N \right)
\end{equation}
where
\begin{align*}
 A_i =  \int_0^1 \Big(   \sum_{t=1}^T \left(w_{b}(t,u) - \bar w_{b}(t)\right)
  e_{i t} \Big)^2 du \\
\textrm{and} \quad
 R_N =N \int_0^1 \Big(  \sum_{t=1}^T  \left( w_{b}(t,u)- \bar w_{b}(t)  \right)
\bar e_{\cdot t}
\Big)^2 du ,
\end{align*}
and we show that asymptotically,  $\sum_i A_i$ is normally
distributed and $R_N$ is negligible.

First, define
\begin{eqnarray*}
 A_i^\circ =  \int_0^1 \Big( \sum_{t=1}^T w_{b}(t,u) e_{i t} \Big)^2
 d u.
\end{eqnarray*}
By Theorem 1 in Zhang and Wu (2011), under the bandwidth conditions
$T b^{3/2} \to \infty$ and $b\to 0$ and the short-range dependence
condition (\ref{eq:srd}), we have
\begin{eqnarray}\label{eq:May11}
Tb^{1/2} \left( A^\circ_i - \E A^\circ_i \right)
 \mathop{\rightarrow}^{\mathcal{L}} N(0, \sigma^2 K_2^*).
\end{eqnarray}
Observing that $A^\circ_1, \ldots, A^\circ_N$ are i.i.d., it
results from (\ref{eq:May11}) and the Lindeberg-Feller CLT that
\begin{equation}\label{eq:May14}
{Tb^{1/2}\over \sqrt N} \sum_{i=1}^N
 \left( A^\circ_i -\E A_i^\circ \right)
 \mathop{\rightarrow}^{\mathcal{L}} N(0, \sigma^2 K_2^*).
\end{equation}

We now show that  $Tb^{1/2}N^{-1/2}\sum_{i=1}^N (A_i^\circ - A_i)$
is negligible as $N,T\to \infty$. Let
\begin{equation*}
J_i = \sum_{t=1}^T \bar w_{b}(t)e_{i t} \:\:  \textrm{and} \:\:
\dot{J}_i = \sum_{t=1}^T \dot{w}_{b}(t) e_{i t},
 \quad \textrm{where} \:\:
\dot{w}_b(t) = \int_0^1 w_{b}(t,u) du   .
\end{equation*}
Noting that $\max_t | \bar w_{b}(t) | = O(T^{-1})$ and $\max_t  |
\dot w_{b}(t) | = O(T^{-1})$, one can obtain from Lemma 1 in Liu
and Wu (2010) that $\|J_i \|_4=O(T^{-1/2})$ and $\| \dot J_i \|_4
= O(T^{-1/2})$. Hence,
\begin{equation}\label{eq:May191}
\left\| A_i^\circ - A_i \right\|_2^2
 = \| J_i^2 - 2 J_i \dot{J}_i \|_2^2 = O(T^{-2} ) 
\end{equation}
and by the i.i.d. character of the $(A_i^\circ - A_i)$, one
deduces that
\begin{equation}\label{neglig ai0-ai}
\bigg\| Tb^{1/2}N^{-1/2}
 \sum_{i=1}^N (A_i^\circ - A_i)\bigg\|_2^2  =O(b).
\end{equation}

We proceed to study the remainder term $(R_N - \E R_N)$ in
\eqref{eq:May12}. By expanding $R_N$ and using the i.i.d.
character of the $N$ time series, one easily finds that
\begin{equation}\label{decomp}
R_N \stackrel{d}{=}
 A_1^\circ +   J_1^2 - 2 J_1 \dot J_1,
\end{equation}
where $ \stackrel{d}{=} $ stands for equality in distribution. The
terms in the above expansion have been studied before. More
precisely, the relations \eqref{eq:May11} and \eqref{eq:May191}
yield
\begin{equation}\label{RN negligible}
\big\| Tb^{1/2}N^{-1/2} \left(R_N - \E R_N\right) \big\|_2^2
  = O(N^{-1}) + O( N^{-1} b ).
\end{equation}

Putting together \eqref{eq:May12}, \eqref{eq:May14}, \eqref{neglig
ai0-ai}, and \eqref{RN negligible}, one obtains the
asymptotic normality \eqref{eq:cltD2}.

\subsubsection{Case (ii): $N$ is fixed}
\label{sec:smallN}
Recall that $e_{it} = G(t/T; \mathcal{F}_{i t})$. For $\zeta_{it}(u) =
G(u; \mathcal{F}_{it})$, define
\begin{eqnarray*}
 \tilde \zeta_{it}(u) = \mathbb{E}
(\zeta_{it}(u)|\varepsilon_{i, t-m+1}, \varepsilon_{i, t-m+2},
\ldots, \varepsilon_{i, t}).
\end{eqnarray*}
Then the process $\{\tilde \zeta_{it}(u)\}_{t \in \Z}$ is
$m$-dependent with long-run variance function $g^\ast$ converging
to $g$ as $m\to\infty$. As in the proof of Theorem 1 in Zhang and Wu (2011), we introduce the martingale difference
\begin{eqnarray*}
 \tilde D^*_{i,t} &=& \sum_{l = 0}^\infty
 \mathbb{E} (\tilde \zeta_{i,t+l}(t/T) | {\cal F}_{i t})
 - \mathbb{E} (\tilde \zeta_{i,t+l}(t/T) | {\cal F}_{i, t-1})\cr
  &=&  \sum_{l = 0}^m
 \mathbb{E} (\tilde \zeta_{i,t+l}(t/T) | {\cal F}_{i t})
 - \mathbb{E} (\tilde \zeta_{i,t+l}(t/T) | {\cal F}_{i, t-1}).
\end{eqnarray*}
Observe that $\tilde D^*_{i,t}$, $1\le t \le T$, are also
$m$-dependent. Let $\tilde D^\dag_{i,t} = \tilde D^*_{i,t} -
\tilde D^*_{\cdot,t}$, where $\tilde D^*_{\cdot,t} = \sum_{i=1}^N
\tilde D^*_{i,t} / N$; let $(\sigma^\ast)^2 = \int_0^1
(g^\ast(u))^2 du$. By the argument of Theorem 1 in Zhang and Wu
(2011), to derive the asymptotic normality $\eqref{eq:cltallN}$,
it suffices to show that as $T\to\infty$,
\begin{eqnarray}\label{eq:Ju252}
\frac{1}{T^2b(N-1)} \sum_{1 \leq t < t' \leq T}
  \Big(K^*\Big(\frac{t - t'}{2Tb}\Big)\Big)^2
  \sum_{i = 1}^N \sum_{i' = 1}^N
  \mathbb{E}(\tilde D^\dag_{i,t} \tilde D^\dag_{i',t})
  \mathbb{E}(\tilde D^\dag_{i,t'} \tilde D^\dag_{i',t'})
  \to K_2^* ( \sigma^\ast) ^2.
\end{eqnarray}

Since the $\tilde D^\dag_{i,t}$, $i=1, \ldots, N$, are i.i.d., we
see that $\mathbb{E}(\tilde D^\dag_{i,t} \tilde D^\dag_{i',t}) =
g^\ast(t) (N-1)/N$ if $i = i'$ and $\mathbb{E}(\tilde D^\dag_{i,t}
\tilde D^\dag_{i',t}) = - g^\ast(t)/N$ if $i \neq i'$. With a few
manipulations, we then obtain
\begin{eqnarray*}
\sum_{i = 1}^N \sum_{i' = 1}^N
 \mathbb{E}(\tilde D^\dag_{i,t} \tilde D^\dag_{i',t})
 \mathbb{E}(\tilde D^\dag_{i,t'} \tilde D^\dag_{i',t'})
 = (N-1) g^\ast(t) g^\ast (t').
\end{eqnarray*}
Furthermore, with the continuity of $g^\ast$,
 classic arguments for kernel smoothing show that
\begin{eqnarray*}
\frac{1}{T^2b} \sum_{1 \leq t < t' \leq T} \Big(K^*\Big(\frac{t -
t'}{2Tb}\Big)\Big)^2 g^\ast(t) g^\ast(t')
 = K_2^* (\sigma^\ast)^2 + o(1).
\end{eqnarray*}
Hence  (\ref{eq:Ju252}) is proved and the asymptotic normality
\eqref{eq:cltallN} follows. $\square$


\subsection{Proof of Theorem \ref{thm:power}}

By \eqref{eq:May12} and the Cauchy-Schwarz inequality, we can
write

\begin{equation}
\widehat{\Delta}_{N,T}
  = I_{N,T} + \Big(\sum_{i=1}^N A_i - R_N\Big)
  + \mathcal{O}_p \left(  I_{N,T}^{1/2}
  \Big( \sum_{i=1}^N A_i - R_N \Big)^{1/2} \right),
\end{equation}
where
\begin{equation*}
I_{N,T} = \sum_{i=1}^N \int_0^1 \Big\{\sum_{t=1}^T\left(w_b(t,u)
 - \bar{w}_b(t)\right)\left(\mu_i(t/T) - \mu(t/T)\right)\Big\}^2 d u
\end{equation*}
and $\mu= N^{-1} \sum_{i=1}^N \mu_i$. By the approximation
properties of local linear smoothers (see for example Proposition
1.13 in p.39 of Tsybakov (2009)), we obtain
\begin{equation}
I_{N,T} = \Delta_N + \mathcal{O}\left(N(b^2 + T^{-1})\right),
\end{equation}
provided that the $\mu_i,i=1,\ldots,N,$ have uniformly bounded
second derivatives on $[0,1]$.

On the other hand, we know from  \eqref{eq:May14}, \eqref{neglig
ai0-ai}, and \eqref{RN negligible} that
\begin{equation}
 Tb^{1/2}N^{-1/2} \Big(  \sum_{i=1}^N
 \left(A_i -\E A_i \right) - R_N \Big)
 \mathop{\rightarrow}^{\mathcal{L}} N(0, \sigma^2 K_2^*).
\end{equation}
By the stochastic Lipschitz continuity \eqref{eq:Jun953} and the
short-range dependence condition \eqref{eq:srd}, and by properties
of weight functions of local linear smoothers (see Lemma 1.3 in
p.38 of Tsybakov (2009)) we also have $\E A_i = \mathcal{O}
((Tb)^{-1})$. Hence,
\begin{equation}
  \widehat{\Delta}_{N,T}
  =  \Delta_N + \mathcal{O}\left(Nb^2\right)
  + \mathcal{O} \left( N/Tb\right) + o_P\left(  N/Tb \right) .
\end{equation}
If $N^{-1}\Delta_N$ converges to 0 at a rate slower than $b^2 +
1/Tb$, then $N^{-1}(b^2 + 1/Tb) \hat{\Delta}_{N,T} \to \infty$ in
probability. Hence the test based on $\hat{\Delta}_{N,T}$ has unit
asymptotic power. $\square$


\subsection{Proof of Theorem \ref{thm:ghat}}

Let $\hat g_i(u) = \sum_{k = -K_T}^{K_T} \hat \gamma_{ik}(u)$ be
the estimated long-run variance based on $\{e_{it}\}_{t=1}^T$, where
$K_T = \lfloor T\tau\varrho \rfloor$ is the truncation order and
$\hat \gamma_{ik} = \frac{1}{|\mathcal{N}_\tau(u)| - |k|} \sum_{
t,t+k  \in  \mathcal{N}_\tau(u) } e_{it} e_{i,t+k}$ is the  sample
autocovariance at lag $k$. Since the cardinality $|N_{\tau}(u) | $ is of order
$T\tau$, one sees that
\begin{equation}\label{approx hat g_i}
\hat g_i(u) = \frac{1+ \mathcal{O}(\varrho)}{ |\mathcal{N}_\tau(u)|}
 \sum_{t \in \mathcal{N}_\tau(u)}
 \sum_{t' \in \mathcal{N}_\tau(u)} e_{it} e_{it'}
 \1I_{\{ |t - t'|  \leq K_T \}} \, .
\end{equation}
By 
the argument of Proposition 1 in Liu and Wu (2010), it can be
shown that $\sup_{u \in \left[0,1\right]}\|\hat g_i(u) -
\mathbb{E} \hat g_i(u)\|_{2} = \mathcal{O}(\sqrt{\varrho})$ and by
the i.i.d. property of the $\{e_{it}\}_{t=1}^T$, one deduces that
\begin{equation}
\sup_{u \in \left[0,1\right]}\|\hat g(u) -
\mathbb{E} \hat g(u)\|_{2} = \mathcal{O}(\sqrt{\varrho/N}).
\end{equation}

The expectation $\E \hat g_i(u) $ can be used to
approximate the truncation of $g(u)$ to order $K_T$ thanks to
the stochastic Lipschitz continuity \eqref{eq:Jun953} and the
martingale decomposition of Wu (2007).
Specifically, let $\Gamma_{2}(k) = \sum_{j=0}^\infty \delta_{2}(j)
\delta_2(j+ k)$. Then for all $u\in [0,1]$ and $t,t'\in
\mathcal{N}_\tau(u)$ such that $|t - t'| \leq K_T $, it holds that
\begin{equation}\label{approx Ee_ite_it'}
\left| \mathbb{E} (e_{it} e_{it'})
- \gamma_{|t-t'|}(t/T) \right|
 \leq C (\Gamma_2(|t-t'|) \wedge (\tau \varrho)).
\end{equation}
Moreover, we obtain after easy calculation that
\begin{equation}\label{maj sum min gamma2 taurho}
\sum_{k=0}^\infty (\Gamma_{2}(k) \wedge (\tau \varrho))
 = \mathcal{O}((\tau \varrho)^{\alpha/(1+\alpha)}).
\end{equation}
Taking the expectation in \eqref{approx hat g_i} and adding terms
so that the summation index set is $\{ (t,t'): t\in
\mathcal{N}_{\tau}(u), 1\le t' \le T, |t-t'| \le K_T \}$, it stems
from  \eqref{approx Ee_ite_it'} and \eqref{maj sum min gamma2
taurho} that
\begin{align}\label{E hat_gi}
\E \hat g_i(u)  & = \frac{1+ \mathcal{O}(\varrho)}
 { |\mathcal{N}_\tau(u)|} \bigg(
 \sum_{t \in \mathcal{N}_\tau(u)}\sum_{t' =1}^T
   \mathbb{E} (e_{it} e_{it'})  \1I_{\{ |t - t'|  \leq K_T \}}
   + \mathcal{O}(K_T) \bigg)\nonumber \\
& =  \frac{1}{ |\mathcal{N}_\tau(u)|}
   \sum_{t\in\mathcal{N}_\tau(u)}
   \sum_{k=-K_T}^{K_T} \E e_{it}e_{i,t+k}
   + \mathcal{O}(\varrho) \nonumber \\
& =  \frac{1}{ |\mathcal{N}_\tau(u)|}
   \sum_{t\in\mathcal{N}_\tau(u)}
   \sum_{k=-K_T}^{K_T} \gamma_k(t/T)
   + \mathcal{O}((\tau \varrho)^{\alpha/(1+\alpha)})
   + \mathcal{O}(\varrho) \nonumber \\
& =  \frac{1}{ |\mathcal{N}_\tau(u)|}
   \sum_{t\in\mathcal{N}_\tau(u)} \Big( g(t/T)
 - 2 \sum_{k=K_T}^{\infty} \gamma_k(t/T) \Big)
 + \mathcal{O}((\tau \varrho)^{\alpha/(1+\alpha)})
  + \mathcal{O}\left(\varrho \right) .
\end{align}

In \eqref{E hat_gi},
a Taylor expansion of $g\in \mathcal{C}^2[0,1 ]$
at order 2 yields
\begin{equation}
\sup_{u \in
[0,1]}\bigg|   \frac{1}{|\mathcal{N}_\tau(u)|}
 \sum_{t \in \mathcal{N}_\tau(u)} g(t/T)
- g(u) \bigg| = \mathcal{O}(\tau^2).
\end{equation}
Also, the martingale decomposition of Wu (2007) can be applied to
show that $\sup_{u\in[0,1]} |\gamma_k(u)| \leq \Gamma_2(k)$, so
that under the assumptions of Theorem 2,
\begin{equation}
\sup_{u\in[0,1]} \sum_{k=K_T}^\infty  |\gamma_k(u)|
 = \mathcal{O}\bigg(  \sum_{k = K_T}^\infty \delta_2(k) \bigg)
 = \mathcal{O}\left( (T\tau\varrho)^{-\alpha} \right).
\end{equation}

Finally, to obtain (\ref{eqn:bndghat}), it suffices to note that
$\mathbb{E} \hat g(u) = \mathbb{E} \hat g_i(u)$.

\bigskip

To derive (\ref{eqn:bndgtilde}), an easy calculation shows that
\begin{eqnarray*}
\tilde g_i(u) - \hat g_i(u)
 & = & {2 \over |\mathcal{N}_\tau(u)|} \sum_{t \in \mathcal{N}_\tau(u)}
 \sum_{t' \in \mathcal{N}_\tau(u)}
  (\hat e_{it} - e_{it}) e_{it'} \1I_{\{|t - t'| \leq K_T \}} \\
 & & + {1 \over |\mathcal{N}_\tau(u)|} \sum_{t \in \mathcal{N}_\tau(u)}
 \sum_{t' \in \mathcal{N}_\tau(u)}
 (\hat e_{it} - e_{it}) (\hat e_{it'} - e_{it'})
 \1I_{\{|t - t'| \leq K_T \}} \cr
 &:=& I_{N,T}(u) + \II_{N,T}(u).
\end{eqnarray*}
Noticing that $\hat e_{it} - e_{it} = \beta_i(t/T) - \sum_{t'' =
1}^T \beta_i(t''/T)w_b(t'',t) - \sum_{t'' = 1}^T e_{it''}
w_b(t'',t)$, we have
\begin{equation*}
\max_{t = 1,\ldots,T} \|\hat e_{it} - e_{it}\|_{p}
 \leq C (b^2 + T^{-1/2}b^{-1/2}).
\end{equation*}

Hence by Lemma 1 in Zhang and Wu (2011), we have $\sup_{u \in
[0,1]} \|I_{N,T}(u)\|_p \leq C \iota$ and $\sup_{u \in [0,1]} \|
\II_{N,T}(u) \|_p \leq C \iota^2$, which proves (\ref{eqn:bndgtilde}).
$\square$

\section*{References}
\par\noindent\hangindent2.3em\hangafter 1
Bickel and Levina (2008). Regularized estimation of large
covariance matrices. \emph{Ann. Statist.} {\bf 36}, 199--227.

\par\noindent\hangindent2.3em\hangafter 1
Bissantz, N., Holzmann, H., and Munk, A. (2005). Testing
parametric assumptions on band- or time-limited signals under noise. \emph{IEEE Trans. Inform. Theory} {\bf 51}, 3796--3805.



\par\noindent\hangindent2.3em\hangafter 1
Degras, D. (2010). Simultaneous confidence bands for nonparametric regression with functional data. \emph{Statist. Sinica}, To appear. Available at http://arxiv.org/abs/0908.1980

\par\noindent\hangindent2.3em\hangafter 1
Delgado, M. A. (1993). Testing the equality of nonparametric regression curves. {\it Stat. Probab. Lett.} {\bf 17}, 199--204.


\par\noindent\hangindent2.3em\hangafter 1
Fan, J. and Lin, S. K. (1998). Test of significance when data are curves. {\it J. Amer. Statist. Assoc.} {\bf 93}, 1007--1021.

\par\noindent\hangindent2.3em\hangafter 1
Fan, J. and Gijbels, I. (1996). \emph{Local Polynomial Modeling and its Applications}. London, U.K.: Chapman \& Hall.

\par\noindent\hangindent2.3em\hangafter 1
Gottschalk, P. G. and Dunn, J. R. (2005). Measuring parallelism, linearity, and relative potency in bioassay and immunoassay data.
{\it J. Biopharm. Statist.} {\bf 15},  437--463.

\par\noindent\hangindent2.3em\hangafter 1
Guo, P. and Oyet, A. J. (2009). On wavelet methods for testing equality of mean response curves. {\it Int. J. Wavelets Multiresolut. Inf. Process.} {\bf 7}, 357--373.

\par\noindent\hangindent2.3em\hangafter 1
Hall, P. and  Hart, J.D. (1990). Bootstrap test for difference between means in nonparametric regression. {\it J. Amer. Statist.
Assoc.} {\bf 85}, 1039--1049.

\par\noindent\hangindent2.3em\hangafter 1
H\"ardle, W. and Marron, J. S. (1990). Semiparametric comparison of regression curves. {\it Ann. Statist.} {\bf 18}, 63--89.

\par\noindent\hangindent2.3em\hangafter 1
H\"ardle, W. and Mammen, E. (1993). Comparing nonparametric versus parametric regression fits. {\it Ann. Statist.} {\bf 21}, 1926--1947.

\par\noindent\hangindent2.3em\hangafter 1
Lavergne, P. (2001). An equality test across nonparametric regressions. {\it J. Econometrics} {\bf 103}, 307--344.

\par\noindent\hangindent2.3em\hangafter 1
Li, F. (2006). Testing for the equality of two nonparametric regression curves with long memory errors. {\it Comm. Statist. Simulation Comput.}
{\bf 35}, 621--643.

\par\noindent\hangindent2.3em\hangafter 1
Liu, W. D. and Wu, W. B. (2010). Asymptotics of spectral density
estimates. {\it Econometric Theory} {\bf 26}, 1218--1245.

\par\noindent\hangindent2.3em\hangafter 1
King, E. C., Hart, J. D. and Wehrly, T. E. (1991). Testing the equality of regression curves using linear smoothers. {\it Statist. Probab. Lett.} {\bf 12}, 239--247.

\par\noindent\hangindent2.3em\hangafter 1
Koul, H. L. and Schick, A. (1997). Testing for the equality of two nonparametric regression curves. {\it J. of Statist. Plann. Inf.} {\bf 65}, 293--314.

\par\noindent\hangindent2.3em\hangafter 1
Munk, A. and H. Dette (1998). Nonparametric comparison of several regression functions: exact and asymptotic theory. {\it Ann. Statist.}, {\bf 6} 2339--2368.

\par\noindent\hangindent2.3em\hangafter 1
Orbe, S., Ferreira, E. and Rodriguez-Poo, J. (2005) Nonparametric
estimation of time varying parameters under shape restrictions.
{\it J. Econometr.}, {\bf 126} 53--77.

\par\noindent\hangindent2.3em\hangafter 1
Park, C., Vaughan, A., Hannig, J. and Kang, K. H. (2009). SiZer
analysis for the comparison of time series. {\it J. Statist.
Plann. Inf.} {\bf 139}, 3974--3988.

\par\noindent\hangindent2.3em\hangafter 1
Pawlak, M. and Stadtm\"uller, U. (2007) Signal Sampling and
Recovery Under Dependent Errors, {\it IEEE Trans. Inf. Theory}
{\bf 53}, 2526--2541.

\par\noindent\hangindent2.3em\hangafter 1
Vilar-Fernandez J. M., Vilar-Fernandez J. A., and Gonzalez-Manteiga W. (2007). Bootstrap tests for nonparametric comparison of regression
curves with dependent errors. {\it Test} {\bf 16}, 123--144.

\par\noindent\hangindent2.3em\hangafter 1
Tsybakov A. B. (2009). {\em Introduction to nonparametric estimation}. Springer, New York.

\par\noindent\hangindent2.3em\hangafter 1
Wu, W. B. (2005). Nonlinear system theory: another look at dependence. {\it Proc. Natl Acad. Sci. USA} {\bf 102}, 14150--14154.

\par\noindent\hangindent2.3em\hangafter 1
Wu, W. B. (2007). Strong invariance principles for dependent random variables. \emph{Ann. Prob.} {\bf 35}, 2294--2320.

\par\noindent\hangindent2.3em\hangafter 1
Wu, W. B. and Zhao, Z. (2007) Inference of Trends in Time Series.
{\it Journal of the Royal Statistical Society, Series B,} {\bf 69}
391--410

\par\noindent\hangindent2.3em\hangafter 1
Wu, W. B. and Zhou, Z. (2011). Gaussian approximations for
non-stationary multiple time series. {\it Statist. Sinica}, To
appear.

\par\noindent\hangindent2.3em\hangafter 1
Young, S. G. and Bowman, A.W. (1995). Non-parametric analysis of
covariance. {\it Biometrics}, {\bf 51} 920--931.

\par\noindent\hangindent2.3em\hangafter 1
Zhang, T. and Wu, W. B. (2011). Testing parametric assumptions of
trends of non-stationary time series. {\it Biometrika}, To appear.

\end{document}